\documentclass[aps,prl,superscriptaddress,twocolumn,twoside,floatfix]{revtex4}

\usepackage{bm}
\usepackage{graphicx}
\usepackage{amsfonts}
\usepackage{amsmath}
\usepackage{amssymb}
\newcommand {\be}{\begin{equation}}
\newcommand {\ee}{\end{equation}}
\newcommand {\bea}{\begin{eqnarray}}
\newcommand {\eea}{\end{eqnarray}}

\begin{document}

\title{Optimal control of the local electromagnetic response of nanostructured materials: optimal detectors and quantum disguises}
\author{Ilya Grigorenko}
\affiliation{Theoretical Division T-1, Center for Nonlinear Studies,
Center for Integrated Nanotechnologies, Los Alamos National
Laboratory, Los Alamos, New Mexico 87545, USA}
\author{Herschel Rabitz}
\affiliation{Chemistry Department, Princeton University, NJ 08544,
USA}
\author{Alaxander Balatsky}
\affiliation{Theoretical Division T-4, Center for Integrated
Nanotechnologies, Los Alamos National Laboratory, Los Alamos, New
Mexico 87545, USA}
\date{\today}

\begin{abstract}
We consider the problem of optimization of an effective trapping
potential in a nanostructure with a quasi-one-dimensional geometry.
The optimization is performed
 to achieve certain target optical properties of
the system. We formulate and solve the optimization problem for a
nanostructure that serves either as a single molecule detector or as
a ``quantum disguise''  for a single molecule.
\end{abstract}

\maketitle

The rapid advance in fabrication techniques  makes it possible to
control the composition and geometry of nanostructured systems and
materials on atomic scales  \cite{hari}. One of the most interesting
and challenging problems in modern nanoscience is how to utilize
these fabrication capabilities to control the interaction of
electromagnetic fields with nanostructured media. A successful
solution to this problem would make possible the design of, for
example, high  energy density storage applications, or
multifunctional nanophotonic materials with predefined properties
\cite{ratner08}.

There is, in general, a complex relationship between the effective
electron trapping potential in a nanostructure and the
nanostructure's optical properties. One can usually solve this
problem for a chosen geometry and structure of the system, thus for
a given trapping potential. However, the inverse problem to find an
optimal geometry of the nanostructure to attain desired optical
properties seems much more complicated. In particular, the
matter-field interaction becomes complex in the quantum limit of
very small nanostructures, containing only a few electrons. Quantum
mechanical corrections and the non-locality of the electromagnetic
response become significant for system sizes comparable to a typical
electron wavelength.
 This condition may be satisfied for such small objects as
quantum dots, or for the next generation of microchips
\cite{levi08}. For example, for the semiconductor $GaAs$ with a
doping level of $10^{18}$cm$^{-3}$ and effective electron mass
$m_e^*=0.07\times m_e$, the Fermi wavelength $\lambda_F\approx20$nm
is on the length scale currently available in microchip design.
Thus, the problem has both fundamental and applied significance, and
there is a strong demand for methods to efficiently explore the
optimal design of quantum nanoscale devices.

A systematic approach to the theoretical analysis of optimal design
of potentials in quantum scattering in semiconductor nanodevices was
first developed in \cite{rabitz94}. Twelve years later,
 an optimal design problem with similar objectives was
considered in \cite{levi06}. In \cite{haas05} a simple tight-binding
model was considered in two dimensions with the aim of fitting a
target density of states, and as a result, also describing the basic
response properties of the system. Another interesting  example of
inverse band structure engineering was considered for nitrogen
impurities in GaP \cite{zunger06}.

One of the potential applications of optimal design of
nanostructures is to the engineering of efficient single molecule
detectors. Single molecule sensing by means of optical excitation is
receiving increasing attention \cite{detector}. In this work we will
demonstrate that with the help of quantum design it is possible to
control and manipulate induced local fields in the nanostructure on
a very fine scale. We formulate and solve the optimal design problem
for nanostructures, which may serve as single molecule detectors. We
consider electrons trapped in the effective potential of the
nanostructure.  The presence of a polar molecule in close vicinity
of the nanostructure may slightly modify the electron trapping
potential, and as a result, its optical response. If  a
monochromatic electromagnetic field of frequency $\omega$ is used to
excite the nanostructure followed by a measurement of the molecular
induced changes, then it is possible to use this nanostructure (or,
better, an array of identical nanostructures) as a single molecule
detector. The optimal design goal is to find a nanostructure with an
effective trapping potential that results in maximum contrast
between the optical response either with or without a particular
molecule at a given excitation frequency. Note, that the opposite
optimization problem, to find the effective trapping potential that
minimizes the contrast, can also be formulated. We also consider
this problem, and refer to this type of nanostructure as providing
{\it quantum disguise}.


 We assume that there are $N_{el}$ electrons trapped in a nanostructure. The system is described by the Hamiltonian
 $H_n=\nabla^2+V_{trap}({\bf r})$, with the eigenenergies and eigenfunctions  $E_i$ and $\psi_i$,
 respectively.
 Since inhomogeneous systems with a few electrons can have an electromagnetic response very different from the bulk,
in our simulations we use the non-local density-density response
function within the coordinate-space representation $\chi ({\bf
r},{\bf r'},\omega )$ :
 \bea \label{chi}\chi ({\bf
r},{\bf r'},\omega ) = \sum\limits_{i,j} \frac{f(E_i ) - f(E_j
)}{E_i - E_j  - \hbar \omega  - i\gamma } \times \nonumber\\ \psi
_i^* ({\bf r}) \psi _i ({\bf r'})\psi _j^* ({\bf r'})\psi _j ({\bf
r}), \eea
 where $f(E_i)$ is the Fermi filling factor and the small constant
$\gamma$ describes level broadening. $\omega$ is the frequency of
the external electromagnetic field. The electron eigenenergies $E_i$
and eigenfunctions $\psi_i$ are obtained numerically using
real-space discretization of the Schr\"odinger equation. Using
Eq.(\ref{chi}) we calculate the induced electric field ${\bf
E}_{ind}({\bf r},\omega)$ in the system within linear response
theory \cite{ilya06}.


The effect of the presence of a molecule is modeled by a potential
which is added to the effective trapping  potential in the
nanostructure $V_{mol}({\bf r})=-A\exp(-B |{\bf r}-{\bf r}_0|^2)$,
where ${\bf r}_0$ is the position of the molecule. The Gaussian
potential is chosen just for demonstration purposes. We also
performed simulations using a $r^{-6}$ potential with a cut-off, and
obtained qualitatively similar results. A more realistic molecular
potential could be deduced from {\it ab initio} simulations.


As stated above, the presence of a molecule can modify the response
of the nanostructure. To characterize the difference between state
$1$ without the molecule  and state $2$ with the molecule attached,
we introduce the contrast ratio between the two responses: \bea
\label{metric} I_{contr}= \int_V ||{\bf E}^1_{ind}({\bf
r},\omega)-{\bf E}^2_{ind}({\bf r},\omega)||^2 d^3{\bf r}/I_0,
 \eea
 where the normalization intensity $I_0=\int_V
||{\bf E}^1_{ind}({\bf r},\omega)||^2 d^3{\bf r}$, and ${\bf
E}^1_{ind}({\bf r},\omega)$, ${\bf E}^2_{ind}({\bf r},\omega)$ are
the induced electric fields in state $1$ and state $2$,
correspondingly. The choice of the normalization intensity $I_{0}$
is dictated by the aim of finding a nanostructure, whose response in
the state $2$ generates a local electric field with the largest
intensity.

We assume that the effective trapping potential $V_{trap}({\bf r})$
can be represented as a sum of $N_p$ model potentials $V_{mod}({\bf
r})$ located at ${\bf r}_i$, $i=1,..,N_p$, which can be controlled
during the material fabrication procedure with sufficient precision:
\begin{equation}
\label{trap_potential} V_{trap}({\bf r})=\sum_{i=1}^{N_p}
V_{mod}({\bf r}-{\bf r}_i).
\end{equation}
For simplicity  we assume $V_{mod}({\bf r})$ to be the Coulomb
potential of a charge $Q$  with a cutoff $a$:
\begin{eqnarray}
\label{model_potential} V_{mod}({\bf r})&=&Q/|{\bf r}|, \;\; |{\bf r}|>a,\nonumber\\
&=&Q/a, \;\; |{\bf r}|\le a.
\end{eqnarray}
 The optimization algorithm  searches for the optimal parameters ${\bf r}_i$ to
maximize or minimize the contrast ratio given by Eq.(\ref{metric}).
The search for the optimal potential $V_{trap}({\bf r})$ is
performed using Brent's ``principal axis'' optimization algorithm
\cite{brent} for a scalar function of several variables.

In order to reduce the computational complexity of the problem we
perform the simulations considering a quasi-1D nanostructure
geometry. We assume that spatial localization of electrons in the
other two dimensions ($y,z$) creates relatively large energy gaps in
the sub-bands, so we can neglect excitation in these directions. The
incoming field is linearly polarized, with the polarization along
the $x$ axis.  Such systems can be created even on an atomic scale
using STM techniques \cite{ho} by placing atoms of different species
in a line. Let $L$ be the length of the system. Then
$E_0=\frac{\hbar^2}{2 m_e L^2}$ is the unit of energy determined by
the geometry of the system. Here $m_e$ is the electron mass. We
perform optimization of the trapping potential at the frequency of
the external field $\hbar \omega=0.1 E_0$. We assume there are
$N_{el}=15$ electrons trapped within the nanostructure. We set the
temperature to $T=0$K, and the level broadening $\gamma=10^{-2}E_0$.
For the model potential in Eq.(\ref{model_potential}) we use the cut
off parameter $a=0.05 L$ and $Q=15 e/N_p$, where $e$ is the electron
charge. For our simulations we set the molecular position at
$x_0=L/2$. We set the parameters of the molecular potential
$V_{mol}({\bf r})$ to $A=0.005 e/L$ and $B=0.14 L^{-2}$.

\begin{figure}
\includegraphics[width=4.cm,bb=100 240 480 545]{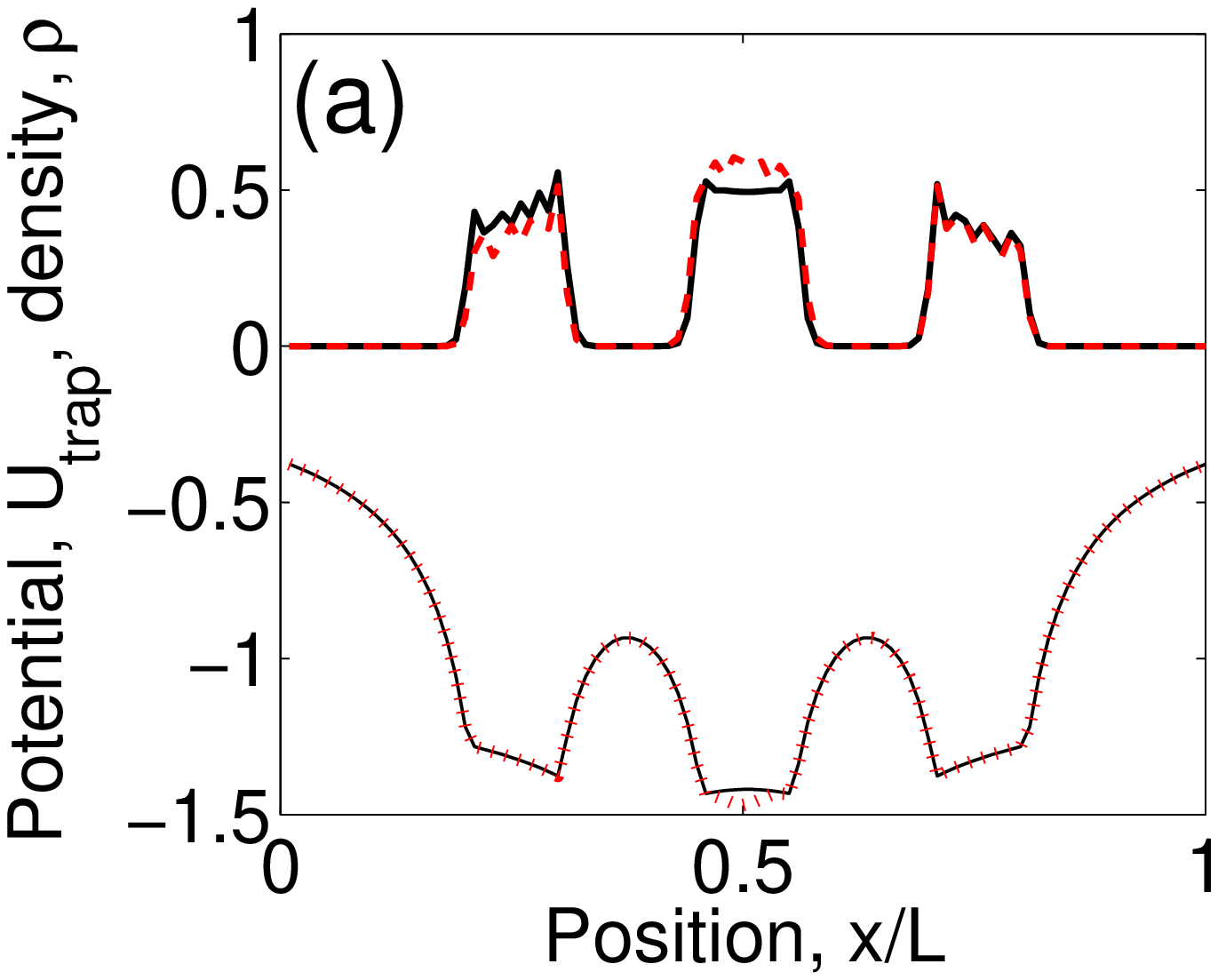}
\includegraphics[width=4.cm,bb=100 240 480 545]{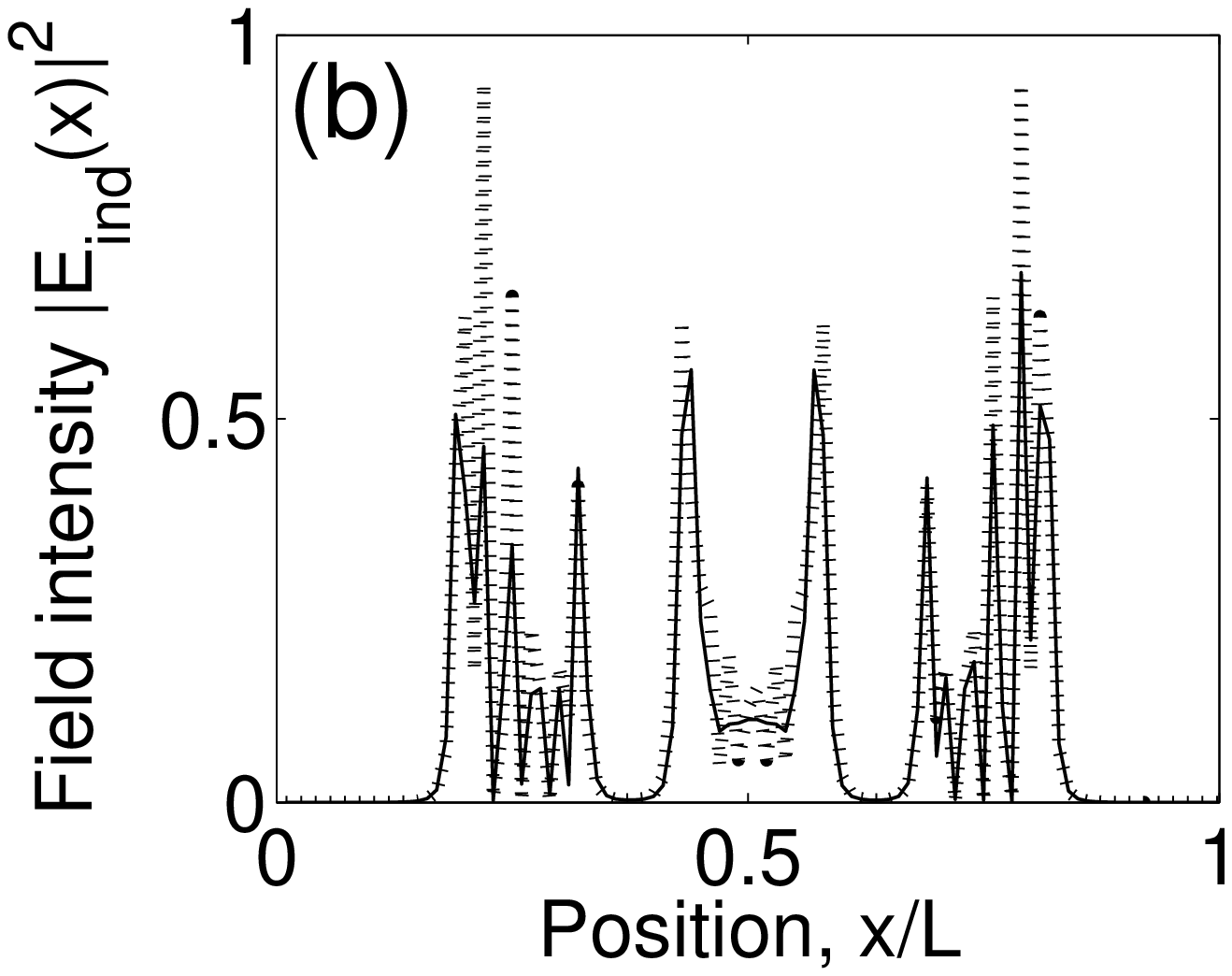}
\includegraphics[width=4.cm,bb=100 240 480 545]{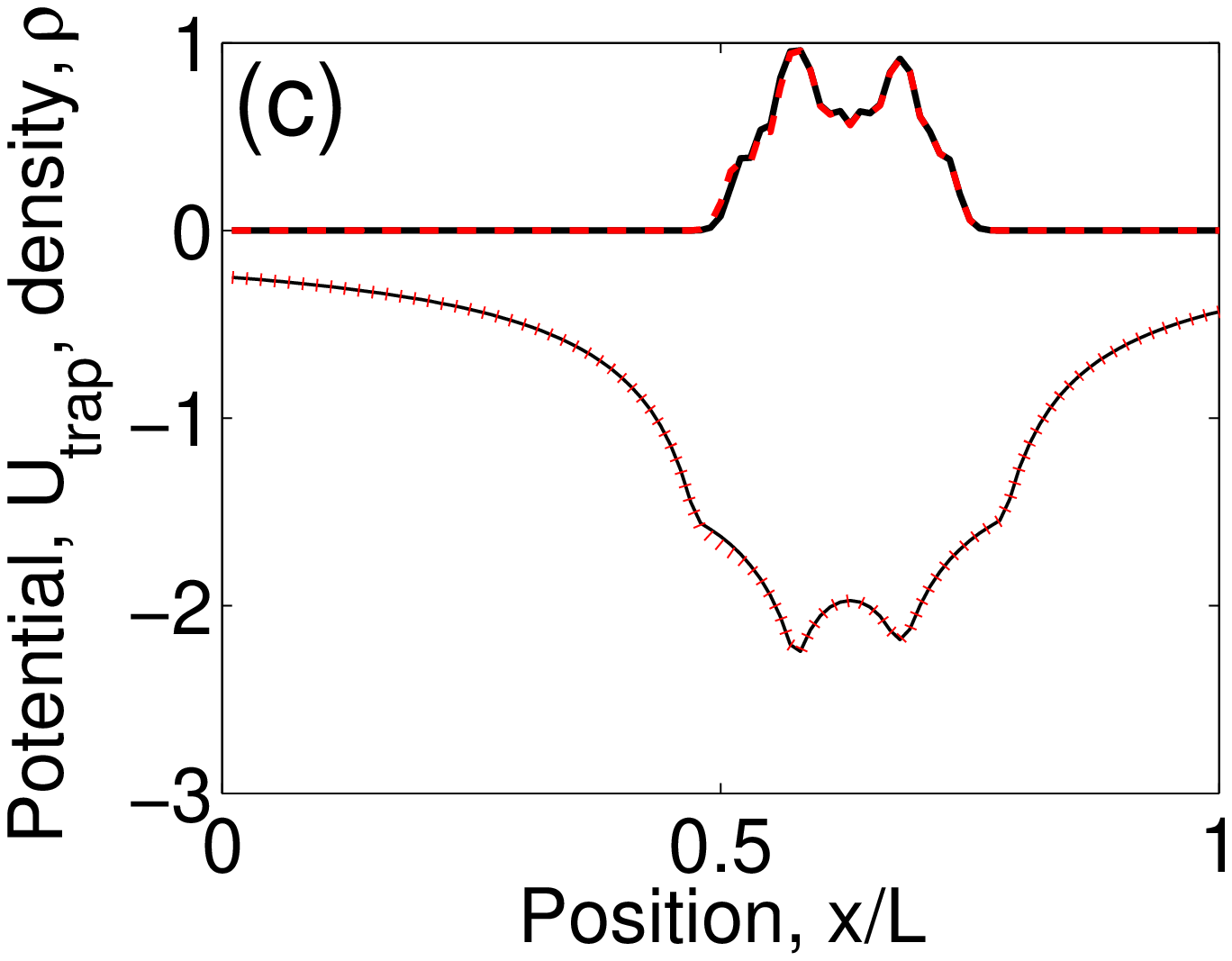}
\includegraphics[width=4.cm,bb=100 240 480 545]{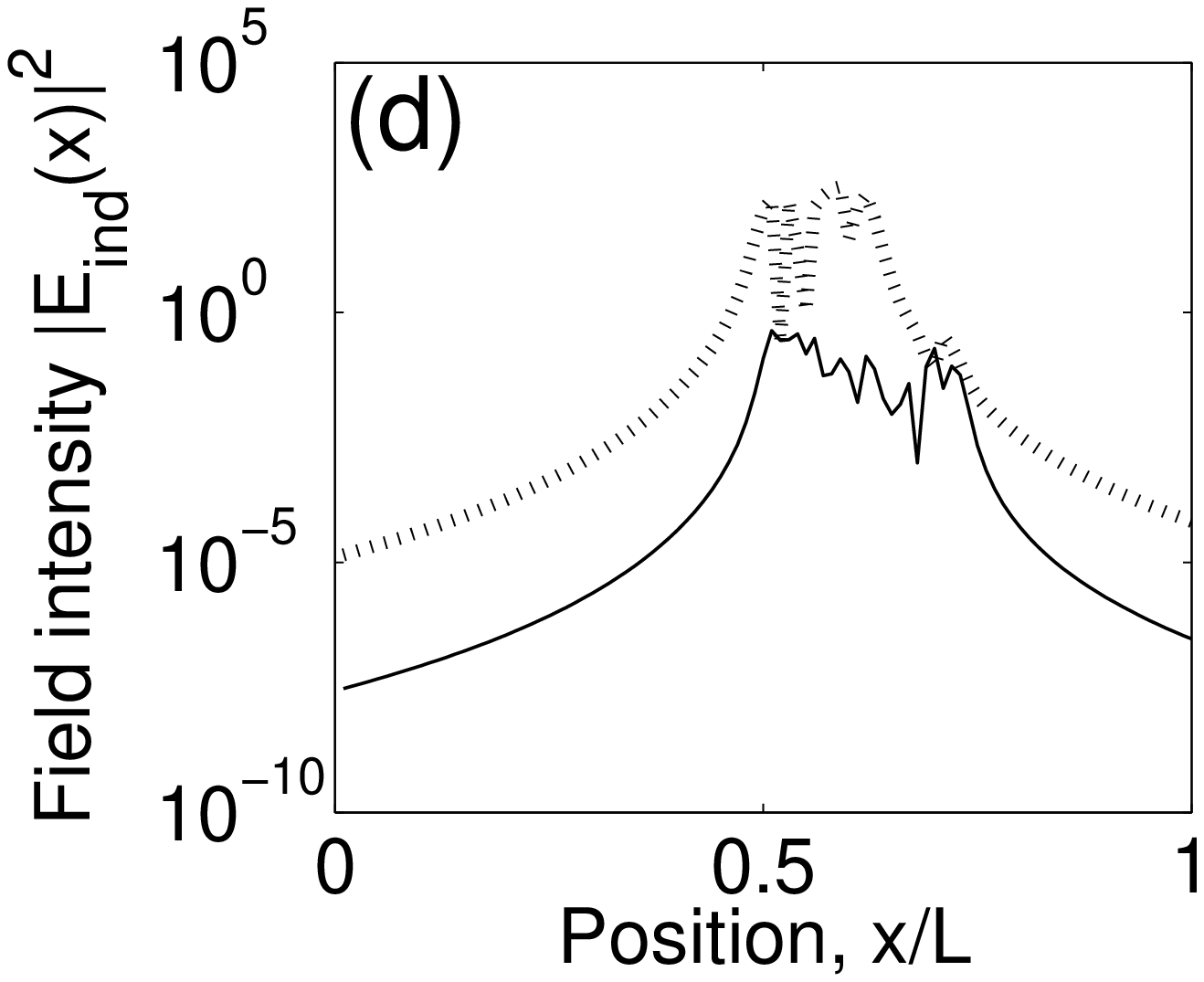}
\caption{\label{fig1}(a)-(b): Initial non-optimized trapping
potential. (a) Trapping potential,  no molecule (thin solid line)
and with molecule (red dotted line). The corresponding electron
density, no molecule (thick solid line) and with a molecule (red
dashed line). (b) Induced field intensity,  no molecule, (solid
line), and with a molecule (dotted line). (c)-(d) Same, as (a)-(b),
but for an optimized potential that {\it maximizes} the contrast
ratio $I_{contr}$ at the frequency of the external field
$\hbar\omega=0.1 E_0$. Note for (d) we used a logarithmic scale for
the $y$-axis. }
\end{figure}

 Figs.~\ref{fig1} (a)-(b)  show the results of calculations for
 a non-optimized nanostructure.  Fig.~\ref{fig1}
 (a) gives
the ground state electron density and the trapping potential without
and with a molecule. Note, that in all the figures the trapping
potential is scaled down by  a factor of $10$. Fig. \ref{fig1}(b)
shows the induced field intensity $|{\bf E}_{ind}|^2$ for the
trapping potentials given in Fig.\ref{fig1}(a).
 Note, that the addition of the potential $V_{mol}(x)$ just slightly modifies the induced field
intensity.

Figs. \ref{fig1}(c)-(d) are the same as \ref{fig1}(a)-(b), but for
an {\it optimized} potential. The optimization is performed to
maximize the contrast ratio given by Eq.(\ref{metric}).  Note, that
in the optimized case, the difference in the induced field intensity
with and without the molecule is so big, that we have to employ a
logarithmic scale. The explanation of this effect is that the
potential $V_{trap}+V_{mol}$ produces an energy difference between
the highest occupied and lowest unoccupied states that very well
matches the frequency of the probe field $\hbar \omega=0.1 E_0$.

Now let us consider the opposite problem of searching for the
trapping potential that minimizes the contrast ratio in
Eq.(\ref{metric}). During the search for the quantum disguise
trapping potential, the optimization procedure has a tendency to
converge to a trivial solution, with the optimized parameters $x_i$
diverging from the molecule position $x_0$. In this case, if there
is a negligible probability of electrons occupying $V_{mol}$, the
presence or absence of
 the molecule does not change the response of the nanostructure. To
 avoid such solutions, we introduce a renormalized contrast ratio $I_{contr}/\delta\rho_g^2$, where $ \label{rho} \delta\rho_g= \int_V
|\rho_1({\bf r})-\rho_2({\bf r})|^2 d^3{\bf r}$, where $\rho_1$ and
$\rho_2$ are the electron ground state density with and without a
molecule, respectively.

\begin{figure}
\vspace{0.cm}
\includegraphics[width=4.cm,angle=0,bb=100 240 480 545]{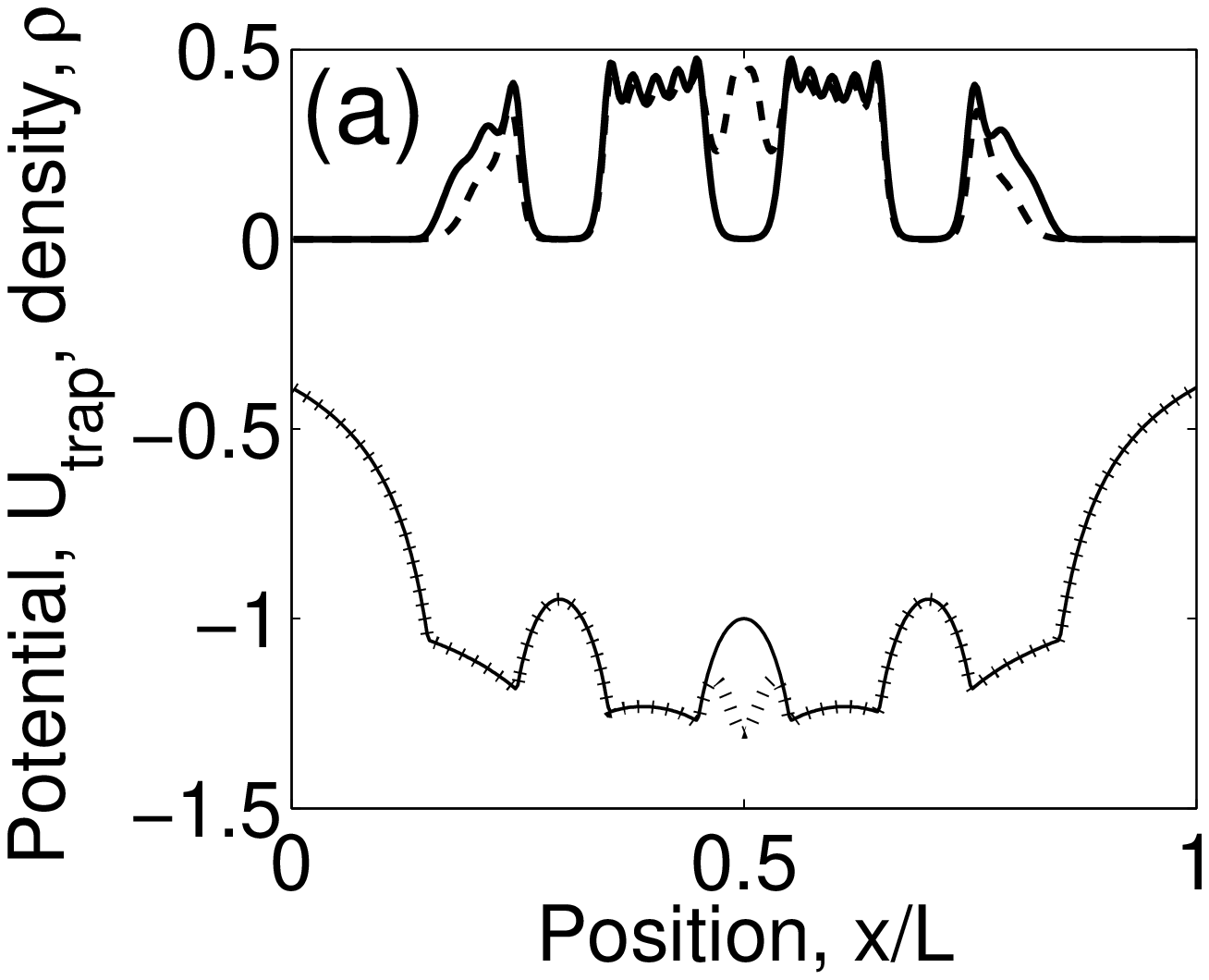}
\includegraphics[width=4.cm,angle=0,bb=100 240 480 545]{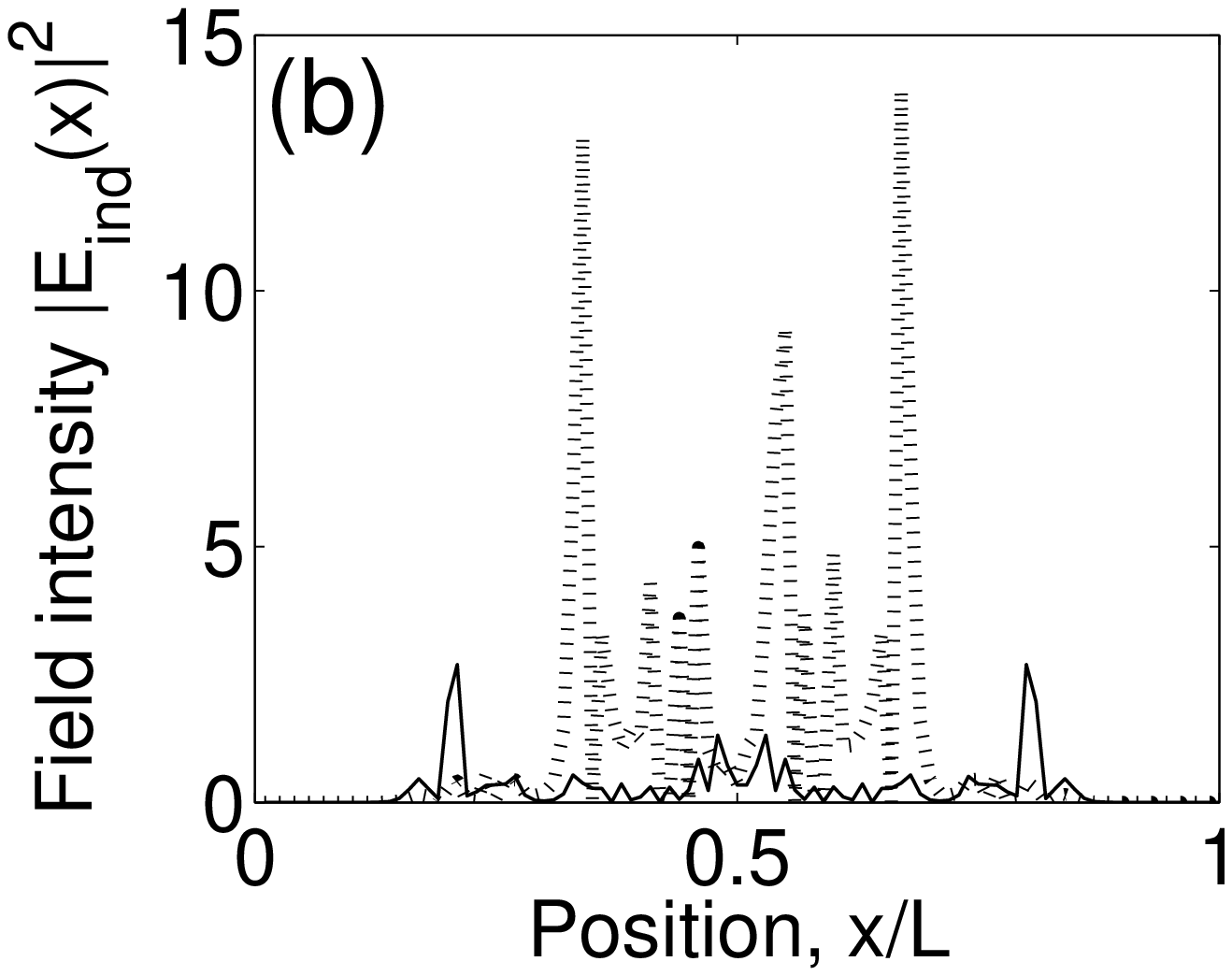}
\includegraphics[width=4.cm,angle=0,bb=100 240 480 545]{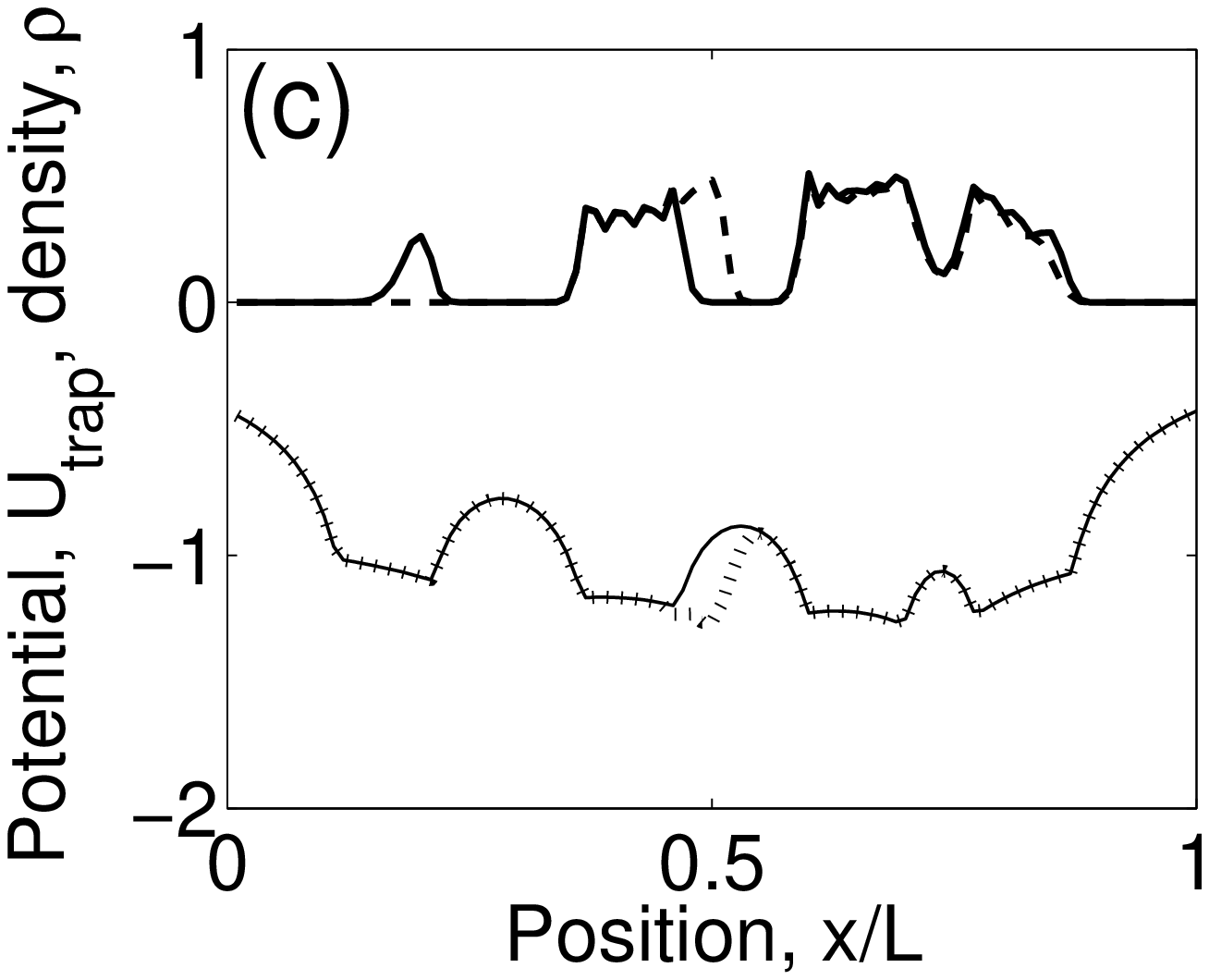}
\includegraphics[width=4.cm,angle=0,bb=100 240 480 545]{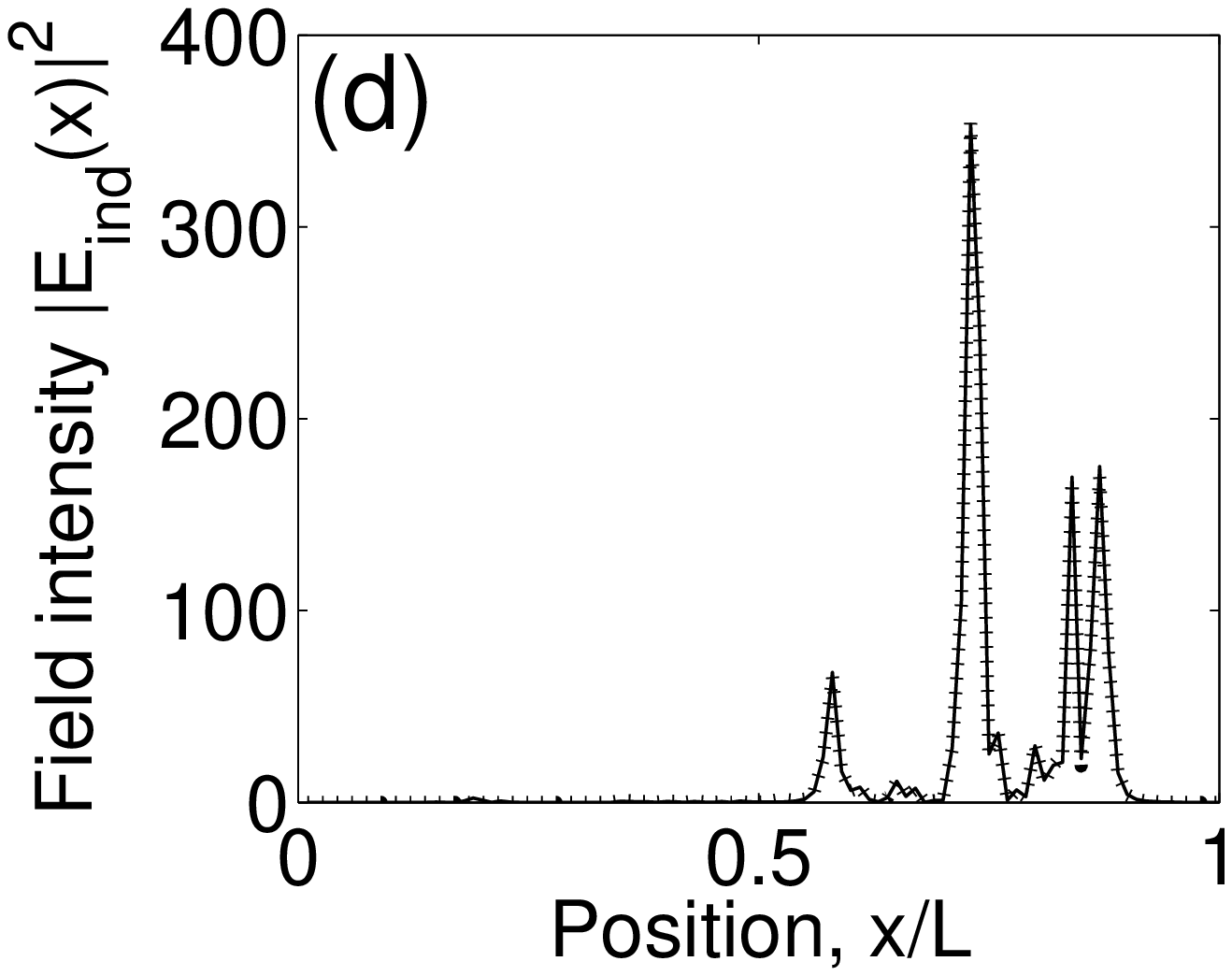}
\caption{\label{fig2} Initial non-optimized trapping potential. (a)
Trapping potential,  no molecule (thin solid line) and with molecule
(dotted line). The corresponding electron density, no molecule
(thick solid line) and with a molecule (dotted line). (b) Induced
field intensity,  no molecule, (solid line), and with a molecule
(dotted line). (c)-(d) Same, as (a)-(b), but for an optimized
potential that {\it minimizes} the contrast ratio $I_{contr}$ at the
frequency of the external field $\hbar\omega=0.1 E_0$. }
\end{figure}

 In Figs.~\ref{fig2} (a)-(c) we present the calculations in the same
 order, as in Figs.~\ref{fig1} (a)-(c), but  for the ``quantum disguise'' potential.
 For the  ``quantum disguise'' problem we set the parameters of the molecular potential
$V_{mol}({\bf r})$ to $A=0.06 e/L$ and $B=0.14 L^{-2}$, so the
perturbation due to the presence of the molecule becomes comparable
to the original trapping potential $V_{trap}$. Note, while the
trapping potentials with and without the molecule, and the
corresponding electron densities shown in Fig.\ref{fig2}(c) are
rather different, the optical responses of both systems at frequency
$\omega=0.1 E_0$ is very close to each other (see
Fig.\ref{fig2}(d)).

The effect of quantum disguise can be explained in terms of quantum
interference between transitions in the combined system
$H_n+V_{mol}$, which cancel each other to give approximately the
response of $H_n$. We checked this assumption by plotting the
induced charge densities $\rho_{ij}(x)$, corresponding to the
transitions between states $i$ and $j$. We found that all these
densities are quite different for the potentials shown in
Fig.\ref{fig2}(a) and Fig.\ref{fig2}(c). However, they all sum up to
almost identical total induced charge densities
$\rho_{ind}(x)=\sum_{i,j}\rho_{ij}(x)$.

\begin{figure}
\includegraphics[width=4.cm,angle=0,bb=100 240 480 545]{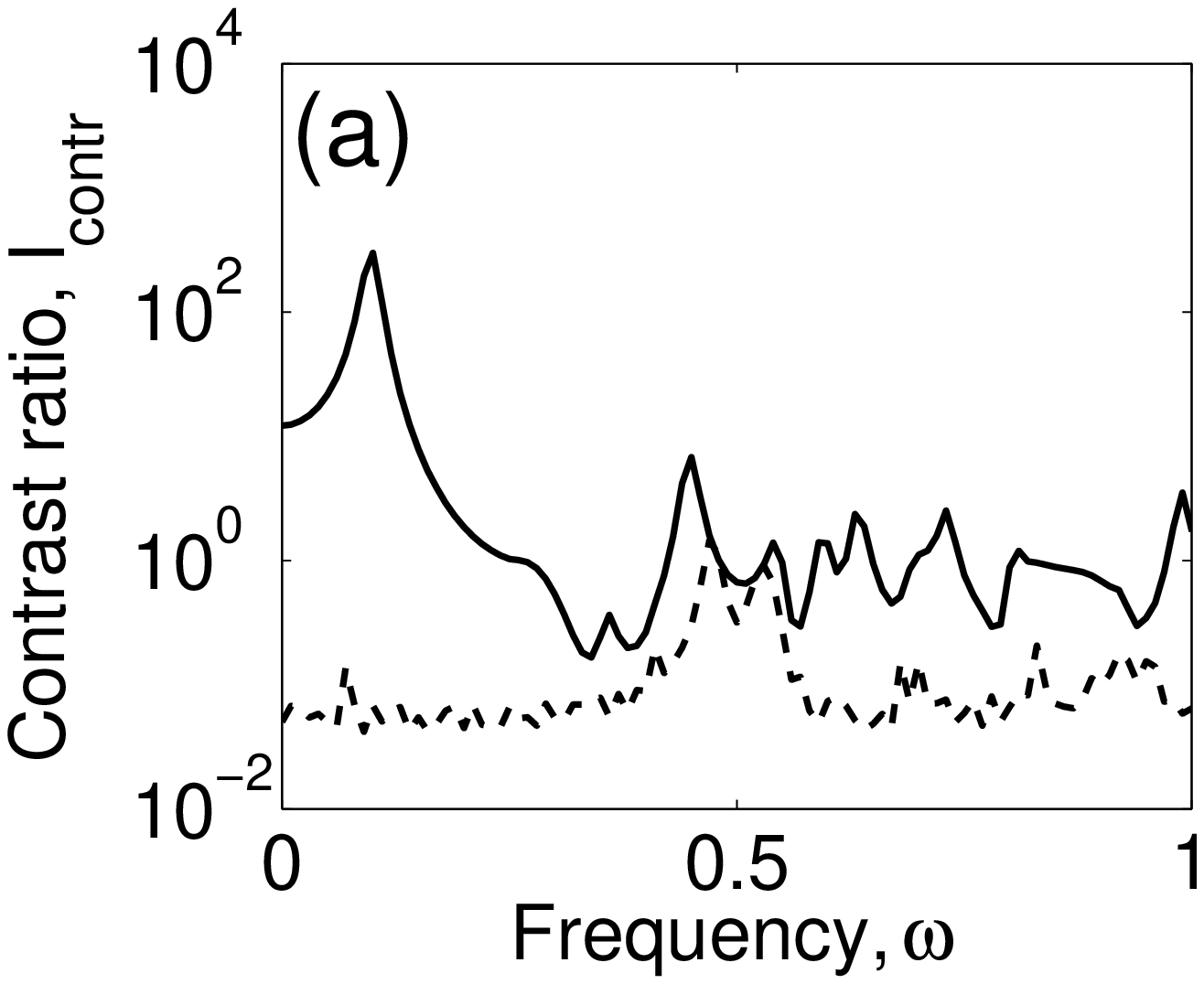}
\includegraphics[width=4.cm,angle=0,bb=100 240 480 545]{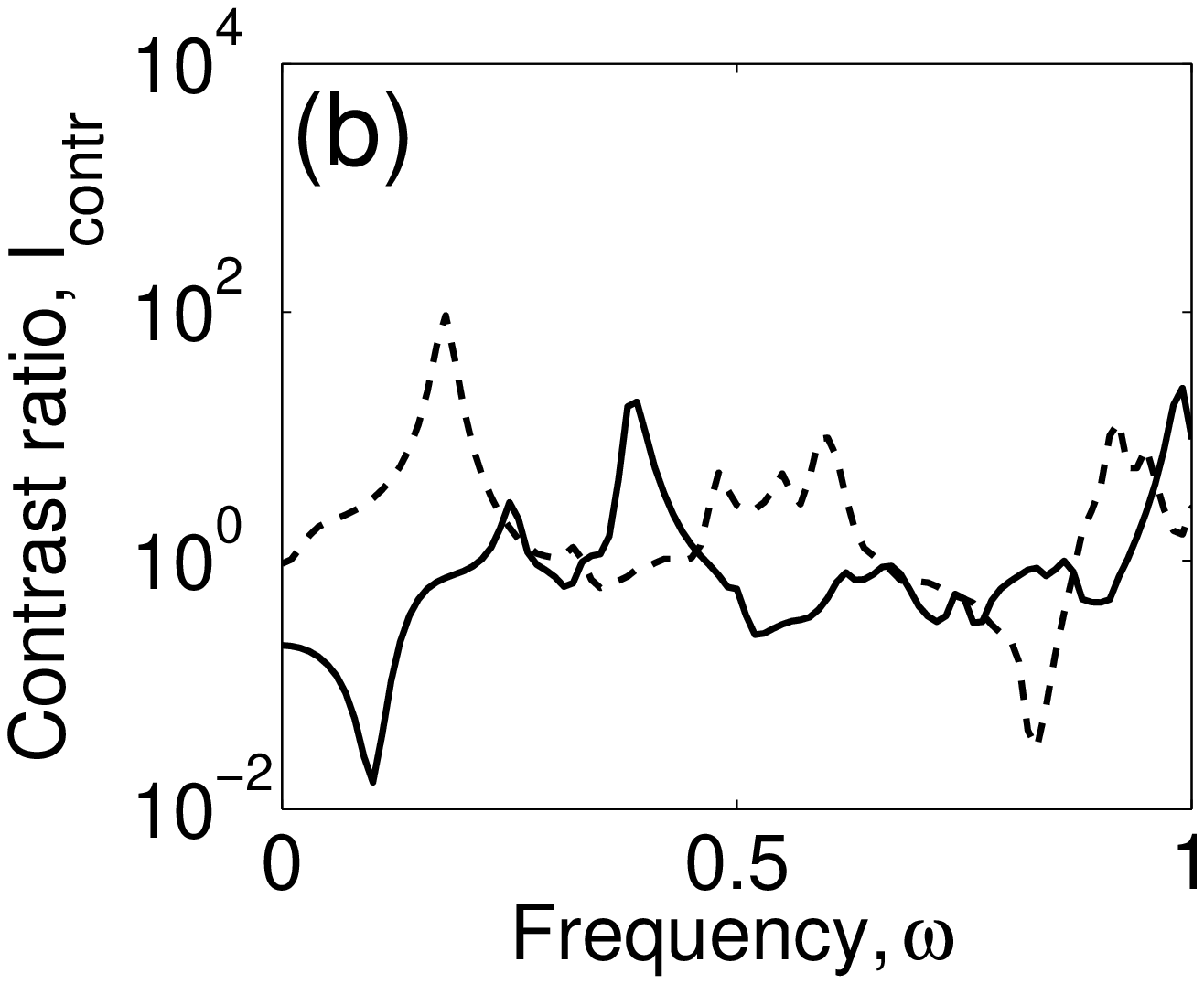}
\caption{\label{fig3}  The contrast ratio $I_{contr}$ between the
induced field intensity, with and without the molecule, as a
function of the frequency of the external field $\omega$. (a) The
dashed line refers to the non-optimized trapping potential shown in
Fig. \ref{fig1}(a), and the solid line is for the optimized
potential in Fig. \ref{fig1}(c). Note the local minimum near the
target frequency $\hbar\omega=0.1 E_0$. (b) Same as (a), but the
contrast ratio is calculated for the potentials shown in Fig.
\ref{fig2}(a) and (c), correspondingly.}
\end{figure}

We also studied the dependence of the nanostructure response  on the
frequency of the external field. Fig. \ref{fig3}(a) shows the
contrast ratio
  as a function of the frequency of the incoming field
$\omega$. The dashed line shows $I_{contr}(\omega)$  for the
initial, non-optimized trapping potential in Fig. \ref{fig1}(a), and
the solid line corresponds to the optimized trapping potential in
Fig. \ref{fig1}(c). Note, that the solid curve has a maximum at the
frequency $\hbar \omega=0.1 E_0$, at which the optimization was
performed. In Fig. \ref{fig3}(b) we plot the contrast ratio
$I_{contr}(\omega)$ for the potentials shown in Figs.
\ref{fig2}(a),(c). Again, the dashed line corresponds to the
initial, non-optimized trapping potential, and the solid line
corresponds to the optimized one. Note the dip near the target
frequency $\hbar \omega=0.1 E_0$, at which the optimization was
performed. In both figures the improvement of the target
functionality $I_{contr}(\omega)$ is of two-three orders of
magnitude.


\begin{figure}
\includegraphics[width=4.cm,angle=0,bb=100 240 480 545]{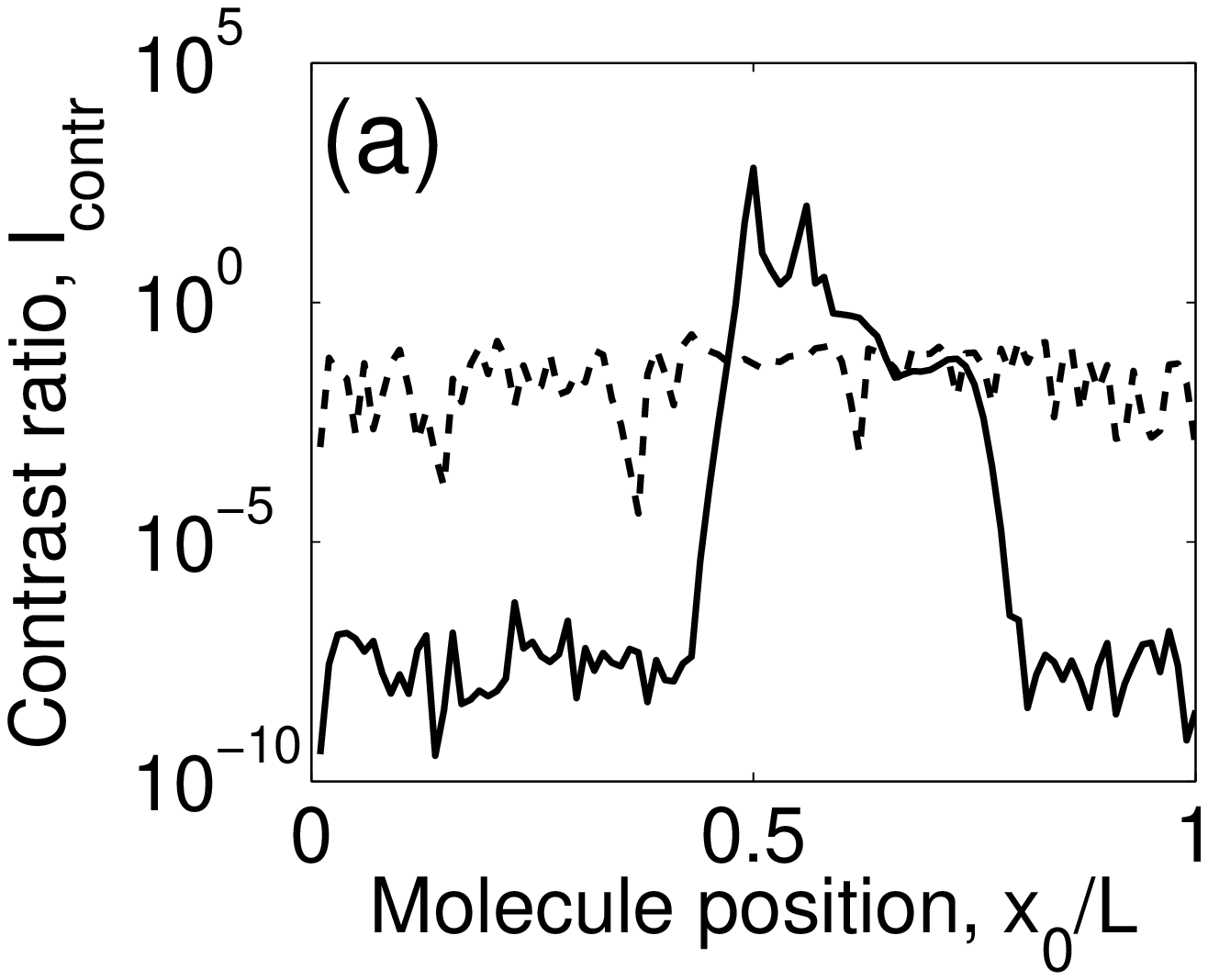}
\includegraphics[width=4.cm,angle=0,bb=100 240 480 545]{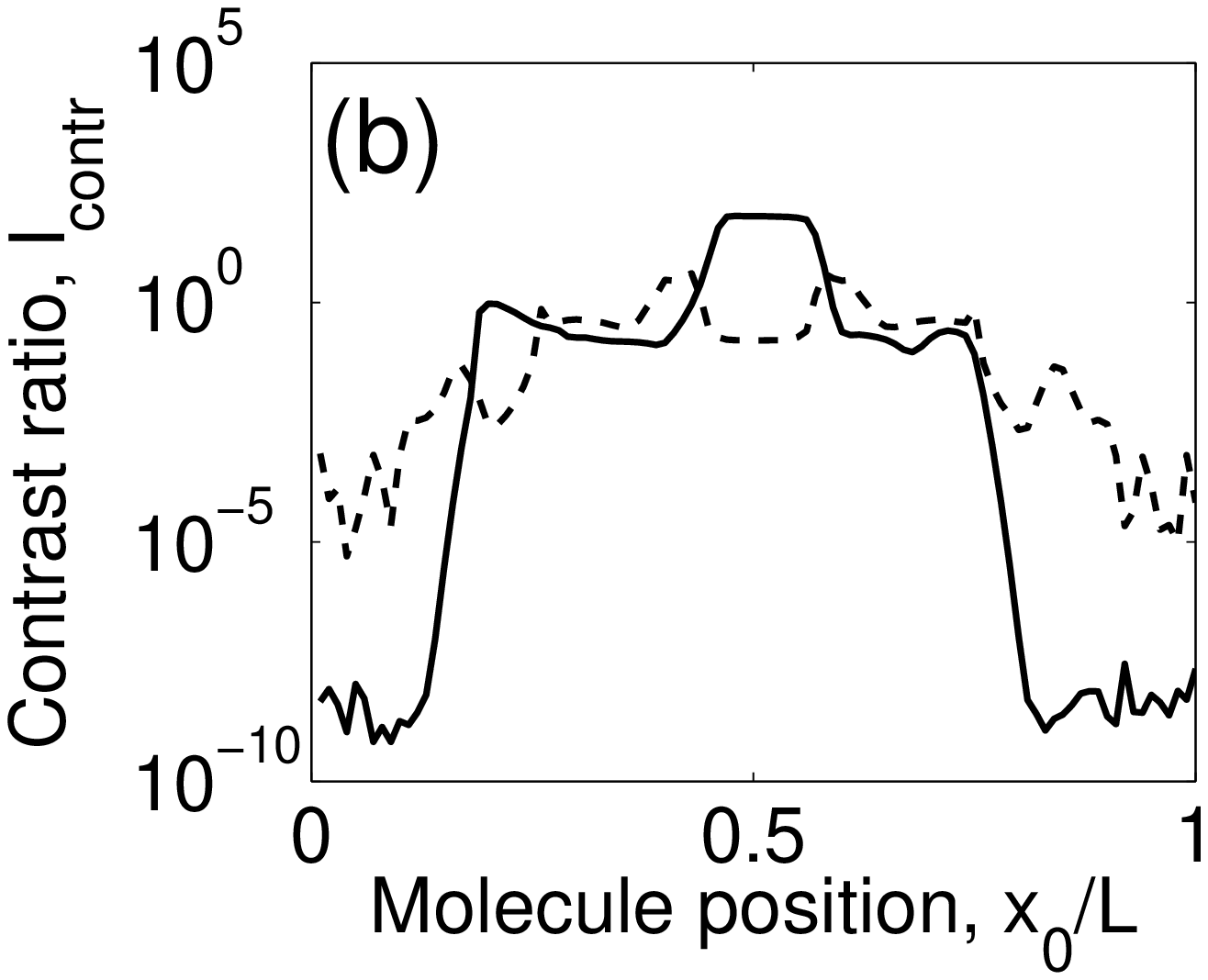}
\caption{\label{fig4}  The contrast ratio $I_{contr}$ between the
induced field intensity, with and without the molecule,  as a
function of position $x_0$ of the molecule. (a) The dashed line
refers to the non-optimized trapping potential shown in Fig.
\ref{fig1}(a), and the solid line is for the optimized potential in
Fig. \ref{fig1}(c).
 (b) Same as (a), but the contrast ratio is calculated for
an optimized potential  using $\{x_i\}$, $i=1,..,5$ parameters.}
\end{figure}
We now consider a molecule position sensitivity analysis. In
Figs.\ref{fig4}(a,b) we show the contrast ratio $I_{contr}$ as a
function of the molecule position for the non-optimized trapping
potential and the optimized one. In  Fig.\ref{fig4}(a) we show the
results for the trapping potential parametrized using $N_{p}=3$, and
in Fig.\ref{fig4}(b) $N_{p}=5$.  It is clear, that even without any
localization mechanism an array consisting of the optimized
nanostructures will outperform a similar array of non-optimized
ones. Indeed, assuming that the molecule's position $x_0$ is
uniformly distributed on $[0,L]$, one can estimate the avarage
enhancement by integrating the area under the curve. The integration
over the curves shown in Fig.\ref{fig4}(a) gives the average value
of the contrast ratio of $S=7.01$ with optimizion,  and $S'=0.46$
for a nonoptimized nanostructure.  Integration of the curves shown
in Fig.\ref{fig4}(b), gives $S=8.43$ and $S'=0.053$ correspondingly.
In the case of molecules localized  near $x_0=L/2$ the enhancement
factor could reach up to three orders of magnitude, compared to the
non-optimized nanostructures.

 In conclusion, we demonstrated the feasibility of optimizing of the trapping potential parameterized by just  a few
control parameters. Even in this limited case the target
functionalities (such as detector sensitivity) can be enhanced by
orders of magnitude.


Since the effects we consider are based on the resonance properties
of the nanostructure,  larger values of the level broadening
$\gamma$ will result in reduced improvement of the design. From a
simple analysis one can estimate that the intensity enhancement for
the optimized design scales as $\propto (\gamma/E_0)^{-2}$. Thus,
phonon, electron-electron scattering and other processes that lead
to level broadening will reduce the optimization effect.


The results of optimal design for the quantum disguise problem show
that although the real and imaginary parts of the nanostructure
response are coupled through the Kramers-Kronig relations and are
not completely independent, we succeeded to make the scattering and
absorbtion properties of two significantly different quantum systems
almost identical at a given frequency .

It is known that for the electron Hamiltonians there is a one-to-one
correspondence (within an additive constant) between the
non-degenerate ground state electron density $\rho({\bf r})$ and the
trapping potential $V_{trap}({\bf r})$. This dependence is
established through the Hohenberg-Kohn mapping \cite{kohn}. Although
some physical properties of the system can be deduced from its
ground state, many characteristics, such as transport properties and
optical response, strongly depend on the excitation spectrum of the
system. The results  suggest that there are situations, when the
precision of the calculated excitation spectrum is crucial. In the
case of the single molecule detector shown in Fig.\ref{fig1}, two
quantum systems with very close ground state densities could have
very different responses. And the results for the quantum disguise
problem show that two quantum systems with almost identical optical
response at a given frequency could have rather different electron
ground states.

The relatively high sensitivity of the optical response of the
optimized nanostructures to fabrication errors can potentially pose
a serious problem. To address this problem we suggest that the
optimal properties of the nanostructures could be controlled during
the assembling phase using closed loop adaptive techniques similar
to those used in optical field control \cite{13}.


This work was performed, in part, at the Center for Integrated
Nanotechnologies, a U.S. Department of Energy, Office of Basic
Energy Sciences user facility.
H.R. also acknowledges support from ARO and NSF.

   \end{document}